\documentclass{article}
\usepackage{graphicx} 
\usepackage[utf8]{inputenc}
\usepackage[
colorlinks=false
]{hyperref}
\usepackage[
backend=biber,
style=alphabetic,
]{biblatex}
\DeclareUnicodeCharacter{2207}{\ensuremath{\nabla}}

\AtEveryBibitem{
    \clearfield{urlyear}
    \clearfield{urlmonth}
    \clearfield{month}
    \clearfield{day}
    \clearfield{url}
}

\usepackage{authblk}
\usepackage{amsmath}
\usepackage{amssymb}
\usepackage[colorinlistoftodos]{todonotes}

\title{The separatrix operational space of next-step fusion experiments: From ASDEX Upgrade data to SPARC scenarios}

\author[1]{T. Eich}
\author[1]{T. Body}
\author[2]{M. Faitsch}
\author[2]{O. Grover}
\author[3]{M. A. Miller}
\author[4]{P. Manz}
\author[1]{T. Looby}
\author[1]{A. Q. Kuang}
\author[2]{A. Redl}
\author[1]{M. Reinke}
\author[1]{A. J. Creely}
\author[1]{D. Battaglia}
\author[1]{J. Hillesheim}
\author[3]{M. Wigram}
\author[3]{J. W. Hughes}
\author{the ASDEX Upgrade team\footnote{See the author list of H. Zohm et al, 2024 Nucl. Fusion https://doi.org/10.1088/1741-4326/ad249d}}

\affil[1]{Commonwealth Fusion Systems, Devens, MA, USA}
\affil[2]{Max-Planck-Institut f\"{u}r Plasmaphysik, Boltzmannstr.2, 85748 Garching, Germany}
\affil[3]{MIT Plasma Science and Fusion Center, MA, USA}
\affil[4]{Institute of Physics, University of Greifswald, 17489 Greifswald, Germany}

\date{Conference paper for PSI 26, Marseille, 2024}

\begin{document}

\maketitle
\section{Abstract}\label{sec:abstract}

Fusion power plants require ELM-free, detached operation to prevent divertor damage and erosion. The separatrix operational space (SepOS) is proposed as a tool for identifying access to the type-I ELM-free quasi-continuous exhaust regime. In this work, we recast the SepOS framework using simple parameters and present dedicated ASDEX Upgrade discharges to demonstrate how to interpret its results. Analyzing an extended ASDEX Upgrade database consisting of 6688 individual measurements, we show that SepOS accurately describes how the H-mode boundary varies with plasma current and magnetic field strength. We then introduce a normalized SepOS framework and LH minimum scaling and show that normalized boundaries across multiple machines are nearly identical, suggesting that the normalized SepOS can be used to translate results between different machines. The LH minimum density predicted by SepOS is found to closely match an experimentally determined multi-machine scaling, which provides a further indirect validation of SepOS across multiple devices. Finally, we demonstrate how SepOS can be used predictively, identifying a viable QCE operational point for SPARC, at $n_{e,sep}=4\times10^{20}m^{-3}$, $T_{e,sep}=156eV$ and $\alpha_t=0.7$ — a value solidly within the QCE operational space on ASDEX Upgrade. This demonstrates how SepOS provides a concise, intuitive method for scoping ELM-free operation on next-step devices.

\section{Introduction}\label{sec:introduction}

Fusion power plants need to be designed with power exhaust in mind to ensure that they achieve the high performance and availability required for commercial success. Tokamak design must address two related challenges — the mitigation of the inter-ELM heat exhaust, and the avoidance or mitigation of large ELMs. Already on existing devices, these challenges require the careful planning of experiments to limit damage to the device walls and pollution of the core plasma. On fusion power plants, however, the challenge will be even more severe and will significantly constrain the viable operational space. Continuous operation with unmitigated type-I ELMs is likely not viable \cite{Zohm2021-lk,Siccinio2020-kl,Kuang2020-za} due to the significant degradation of the thermo-mechanical properties of the divertor targets under high cumulative neutron fluences, and the projected increase in ELM energy fluxes\cite{Eich2017-sh}. Continuous divertor detachment will be required throughout the flattop as well as part of the ramp-up and ramp-down to reduce the risk of tile-cracking due to thermal stresses, melting due to bulk heating, and target erosion and core pollution due to sputtering. Fusion power plants must therefore be able to achieve sufficient performance and power production in detached, ELM-free scenarios — and, ideally, should be designed for access to such scenarios. For this, we need models which are accurate enough to drive the optimization in the correct direction, while still being fast enough to quickly evaluate many potential design points in an optimization loop.

We propose the extended separatrix operational space (SepOS) as a scoping tool, building on the model developed to describe ASDEX Upgrade discharges in Eich \& Manz et al., 2021 \cite{Eich2021-df}. We first present the model in section \ref{sec:sepos} — providing a concise set of easily-implementable equations. We then highlight work to validate the model. To demonstrate how the SepOS describes transitions between different regimes for fixed conditions, we show a series of dedicated ASDEX Upgrade discharges in section \ref{sec:dedicated_discharges}. We then extend this comparison to show that the framework is valid even as the magnetic field strength and plasma current is varied in section \ref{sec:varying_parameters}.

Varying these parameters is shown to shift the regime boundaries. To compare the SepOS predicted for different conditions, we develop a scaling for the LH minimum-density point in section \ref{sec:normalized_sepos} and show that normalizing to this point results in nearly identical operational spaces across different conditions. This finding hints that we can use the framework to translate the results of one device to another. As a demonstration of how the normalized SepOS can be used for scoping ELM-free operations, we apply it to identify the QCE regime on SPARC in section \ref{sec:qce_access_on_sparc}. Finally, we discuss ongoing and future work to validate the SepOS framework under a broader range of conditions, including on other machines and with strong impurity seeding for detachment access. This ongoing work will be used to extend the SepOS framework, towards a reliable tool for designing fusion power plants with tolerable edge conditions in center view.

\section{The separatrix operational space expressed with local parameters}\label{sec:sepos}

\subsection{A data-driven approach}

\begin{figure}
    \centering
    \includegraphics[width=0.7\linewidth]{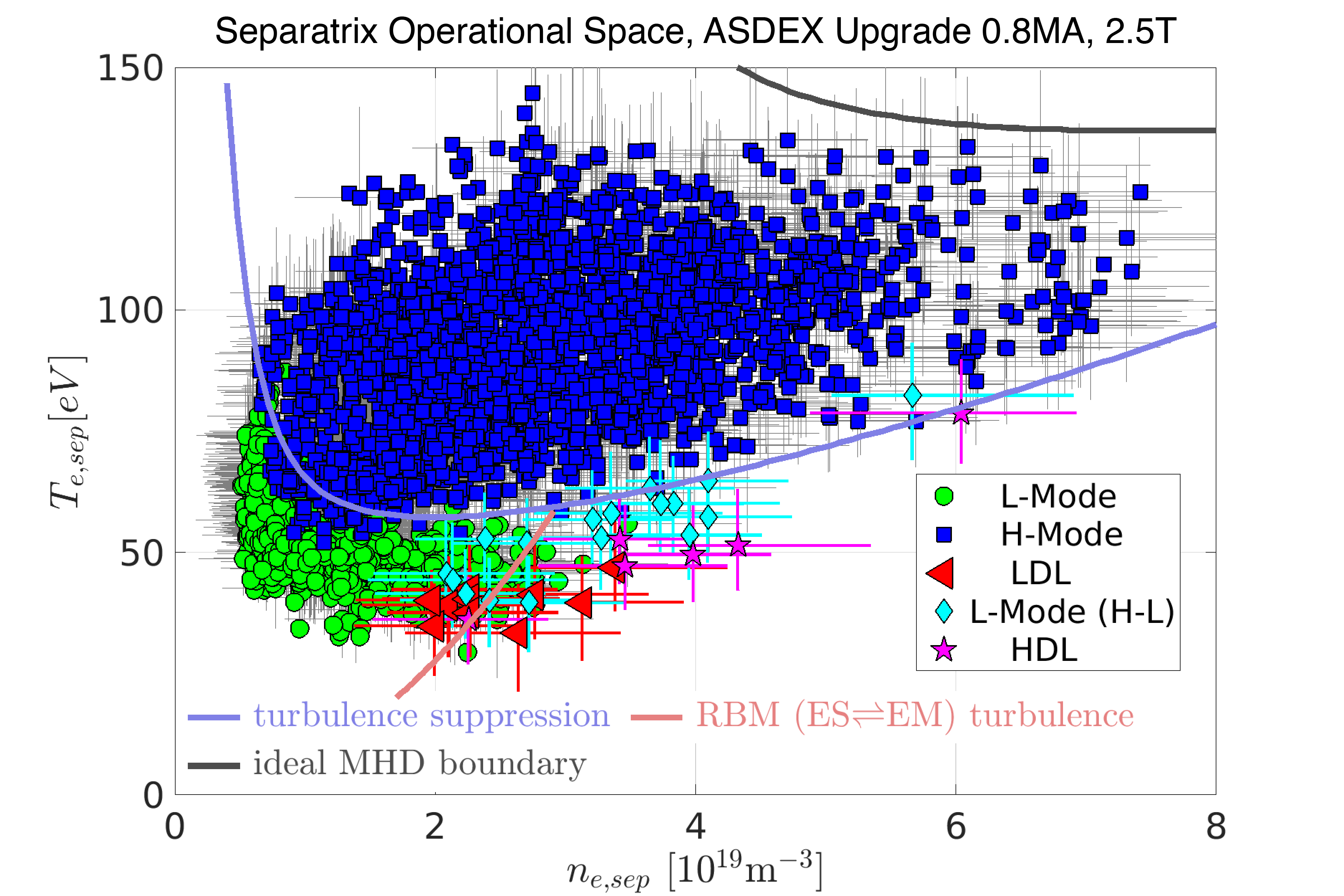}
    \caption{The separatrix operational space of ASDEX Upgrade at $I_p=0.8MA$, $B_t=2.5MA$. Each point gives the separatrix density $n_{e,sep}$ and separatrix temperature $T_{e,sep}$ measured by edge Thomson scattering (following the method in section \ref{sec:spitzerharm}), for multiple time-points within individual discharges and across a large number of discharges (n.b. the experimental database has been expanded significantly since ref \cite{Eich2021-df}). The color and marker of each point indicate whether that point corresponds to a measurement during L-mode (\textit{green circles}), H-mode (\textit{blue squares}), an L-mode density limit (\textit{red triangles}), a HL back transition (\textit{cyan diamonds}) or a H-mode density limit (\textit{magenta stars}). The separatrix operational space was developed to explain the clear separation of the points. This gives the LH-transition (\textit{blue line}, see section \ref{subsec:LH}), the L-mode density limit (\textit{red line}, see section \ref{subsec:LDL}) and the ideal MHD limit (\textit{black line}, see section \ref{subsec:MHD}).}
    \label{fig:sepos_aug_0.8MA_2.5T}
\end{figure}

In ASDEX Upgrade, we can classify discharges based on their separatrix density $n_{e,sep}$ and temperature $T_{e,sep}$ \cite{Eich2021-df}. In figure  \ref{fig:sepos_aug_0.8MA_2.5T}, H-modes (indicated in \textit{blue}) have a higher separatrix temperature than L-modes (indicated in $green$) for a given separatrix density. Above a certain density, density limit disruptions are observed — either as L-mode density limits (\textit{red}), HL back-transitions leading to L-mode density limits (\textit{cyan}) or H-mode density limits where no HL back-transition is observed (\textit{magenta}). To describe the boundaries between L-modes, H-modes and density-limit-disruptions, we introduced the separatrix operational space (SepOS) in Eich \& Manz et al., 2021 \cite{Eich2021-df}. This framework describes the observed boundaries in terms of turbulent growth rates and shear-flow suppression close to the separatrix — building on the work of Rogers, Drake \& Zeiler, 1998\cite{Rogers1998-xp} and Scott, 2005 \cite{Scott2005-ig}. A key advantage of our approach compared to these earlier works was the availability of a large database of edge Thomson-scattering measurements from ASDEX Upgrade, which let us quickly check the validity of our proposed model. As seen in figure \ref{fig:sepos_aug_0.8MA_2.5T}, this data-driven approach reproduces the experimental results remarkably well. However, it also highlights the need for careful validation of the model for other conditions — as we show in section \ref{sec:varying_parameters} — and for other machines and beyond the existing database — as we discuss in section \ref{sec:future_work}.

In this section we present the central equations of the separatrix operational space, expressed in terms of separatrix parameters, gradients and global machine parameters. The end result is a set of easily interpretable equations, which has been implemented in the publicly-available \href{https://github.com/cfs-energy/cfspopcon}{github.com/cfs-energy/cfspopcon}. We do not intend for this section to be a complete discussion of the development or physics of the SepOS framework. Instead, we refer the reader to
\begin{itemize}
\item Rogers, Drake \& Zeiler, 1998 \cite{Rogers1998-xp}, Scott, 2005 \cite{Scott2005-ig} and Chapters 11-14 of Scott, 2021 \cite{Scott2021-ul} for a discussion of the underlying physics,
\item Eich et al., 2020 \cite{Eich2020-jx} for the parametrization of near-separatrix gradients in terms of $\alpha_t$,
\item Eich \& Manz et al., 2021 \cite{Eich2021-df} for the basics of the separatrix operational space,
\item Manz, Eich \& Grover et al., 2023 \cite{Manz2023-vo} for an extended discussion of the L- and H- mode density limit,
\item Faitsch et al., 2023 \cite{Faitsch2023-ws} for the application of the SepOS framework to identify QCE access, and
\item Grover et al., 2024 \cite{Grover2024-hd} for an extension to treat unfavorable $\nabla B$ ion drift cases.
\end{itemize}

\subsection{Identifying the separatrix using Spitzer-Harm power balancing}\label{sec:spitzerharm}

The availability of high-resolution density and electron temperature profiles from the edge Thomson scattering system on ASDEX Upgrade was crucially important for the development of the separatrix operational space. The Thomson scattering measurement does not directly identify the separatrix, however, and due to the steep profiles in the edge, small errors in the magnetically-reconstructed separatrix position can introduce significant errors in the separatrix values. Instead, to determine the position of the separatrix, the separatrix temperature is estimated via Spitzer-Harm power balancing
\begin{align}
    T_{e,sep} \approx \left( \frac{7}{2} \frac{f_{cond} f_{tar}  q_{\parallel,u} L_\parallel}{\kappa_{e,0}} \right)^{2/7} \label{eq:Te_sep}
\end{align}
where $f_{cond}\approx 1$ is the conducted power fraction, $q_{\parallel,u}=\frac{P_{sep}f_{share}}{2\pi (R+a)\lambda_q}\frac{B_{t,omp}}{B_{p,omp}}$ is the upstream parallel heat flux density directed towards the outboard target (for $\lambda_q$ the turbulence-broadened heat-flux decay length \cite{Eich2020-jx}), $L_\parallel$ is the parallel connection length from the outboard midplane to the divertor target and $\kappa_{e,0}$ is the electron heat conductivity constant. We identify the separatrix as the point where the temperature profile measured via Thomson scattering matches the estimated $T_{e,sep}$ — which lets us calculate the separatrix density $n_{e,sep}$. In practice, we actually use a point at $\rho=0.999$, which is taken at a position $0.57mm$ radially inwards at the outboard midplane from the separatrix position determined above.

\subsection[the turbulence parameter]{The turbulence parameter $\alpha_t$}\label{subsec:alphat}

A key parameter in the SepOS framework is the turbulence parameter $\alpha_t$. This was originally introduced in Scott, 2005 \cite{Scott2005-ig} (denoted there as $C \omega_B$), and it can be interpreted as giving the relative strength of interchange turbulence (dominant at $\alpha_t\sim 1$) relative to drift-wave turbulence (dominant at $\alpha_t\sim 0$) \cite{Stagni2022-pw}. We provide here several definitions, directly in equation \ref{eq:alpha_t_full}, in a simplified form which highlights how it depends on the separatrix values in equation \ref{eq:simplified_alpha_t} and in terms of the edge collisionality in equation \ref{eq:alpha_t_nu_edge}.
\begin{align}
    \alpha_t &= 1.02 \frac{\nu_{ei}}{c_s} \hat{q}_{cyl}^2 R \left( 1 + \frac{m_e}{m_i} \right) \left( 1 + \frac{T_{i,sep}}{T_{e,sep}} \frac{1}{\langle Z \rangle}\right) \label{eq:alpha_t_full}\\
    &\approx 3.13 \times 10^{-18} \hat{q}_{cyl}^2 R \frac{n_{e,sep}}{T_{e,sep}^2} Z_{eff,sep} \label{eq:simplified_alpha_t}\\
     &\approx  \frac{1}{100} \nu_{e,edge}^* \hat{q}_{cyl} \label{eq:alpha_t_nu_edge}
\end{align}
where $\nu_{ei}$ is the electron-ion collision frequency in the edge, $c_s$ is the ion sound speed, $\hat{q}_{cyl}$ is the cylindrical safety factor (defined via equation K.6 in ref \cite{Eich2021-df}), $R$ is the major radius, $m_e$ and $m_i$ are the electron and main-ion masses, $T_{e,sep}$ and $T_{i,sep}$ are the electron and ion temperatures at the separatrix, $\langle Z \rangle$ is the mean ion charge and $\nu_{e,edge}^*$ is the edge collisionality. Here, electron and ion temperatures are set equal.

As $\alpha_t$ is increased the interchange drive in the edge increases, increasing the filamentary cross-field transport and increasing the frequency at which scrape-off-layer filaments are born \cite{Faitsch2021-rp,Stagni2022-pw,Redl2023-jx}. Above a certain critical value of $\alpha_t$ and for sufficient shaping ($\delta_{95} \gtrsim0.3$), the pedestal foot becomes ballooning unstable \cite{Radovanovic2022-nb,Harrer2022-ny}. The resulting transport reaches a level sufficient to prevent the pedestal from reaching the peeling-ballooning limit, eliminating large type-I ELMs. This type-I ELM-free regime is called the `quasi-continuous exhaust' (QCE) regime \cite{Faitsch2021-rp}, since the filaments generated in this regime have a much higher frequency and lower amplitude than type-I ELMs. In Faitsch et al., 2023 \cite{Faitsch2023-ws} it was shown that the QCE operational space on ASDEX upgrade is found for points above the LH transition and with $\alpha_t > 0.55$, giving a simple metric for identifying this type-I-ELM free regime.

\subsection{The LHL boundary}\label{subsec:LH}

In Kim \& Diamond, 2003 \cite{Kim2003-pi} it was proposed that a sustained H-mode requires that the rate of energy lost to a shear flow must exceed the rate of energy production due to turbulence\footnote{This was later investigated in Manz et al., 2012 \cite{Manz2012-bp})}. We used this argument to develop a simple condition which, if fulfilled, indicates that an operational point should be in H-mode;
\begin{align}
    \alpha_{RS}\frac{k_{EM}^3}{1+\left( \frac{\alpha_t}{\alpha_c} k_{EM}^2\right)^2} > \frac{\alpha_t}{\alpha_{c}} \left(\frac{1}{2}k_{EM}^2+k_{EM}^4\right)
+\frac{1}{2}\sqrt{\frac{2\lambda_{p,e,H}}{R}} \label{eq:LH}
\end{align}
Here, the left-hand side of the equation represents the turbulent energy lost to the shear flow, and the right-hand side represents the energy produced by turbulence in the electron (first term) and ion (second term) channels. This expression is given in terms of $\alpha_t$ (given by equation \ref{eq:alpha_t_full}), the Reynold-stress factor $\alpha_{RS} = \begin{cases}1 \text{ in forward field}\\0.4 \text{ in reversed field}\end{cases}$  from Grover et al., 2024 \cite{Grover2024-hd}, the critical value for the ballooning drive (originally from ref \cite{Bernard1983-hw});
\begin{align}
    \alpha_c=\kappa_{sep}^{1.2}\left(1 + 1.5\delta_{sep}\right)\label{eq:alpha_c}
\end{align}
the spatial scale at which electromagnetic induction overcomes electron inertia;
\begin{align}
    k_{EM}=\beta_{e,sep} / \frac{m_e}{m_i}=\frac{2\mu_0 n_{e,sep} T_{e,sep} m_i}{B^2 m_e}\label{eq:kEM}
\end{align}
the electron pressure gradient near the separatrix in H-mode $\lambda_{p,e,H}$ (given in terms of the sound Larmor radius $\rho_{s0}$ and $\alpha_t$ in equation K.1 of ref \cite{Eich2021-df}) and the major radius $R$.

From equation \ref{eq:simplified_alpha_t} we see that $\alpha_t$ is large for high densities and low temperatures. Under these conditions, the electron turbulent drive (the first term on the right-hand-side of condition \ref{eq:LH}) dominates, setting the H-mode transition above the LH-minimum density. Below the minimum density, $\alpha_t$ is small and the ion turbulent drive (the second term on the right-hand-side of condition \ref{eq:LH}) dominates.

Expressing the LH transition in terms of a separatrix temperature instead of a power at first seems difficult to reconcile with established scalings such as Martin et al., 2008 \cite{Martin2008-vw}. However, as shown in ref \cite{Eich2021-df}, for a given $T_{e,sep}$, there should be an associated $P_{sep,e}$ to fulfil equation \ref{eq:Te_sep} — which lets us recast the LH transition in more familiar terms. More challenging, however, is the treatment of the energy required in the ion channel. For a given ion temperature gradient and ion cross-field heat diffusivity $\chi_i$, we can calculate the necessary $P_{sep,i}$. However, there are no established scalings for $\chi_i$, which introduces significant uncertainty. As we discuss in section \ref{sec:future_work}, determining such as scaling could be a key application for multi-machine studies and high-fidelity modelling.

\subsection{The L-mode density limit\label{subsec:LDL}}

As $\alpha_t$ is increased, the plasma transport in the edge increases\cite{LaBombard2005-ko}. At intermediate values, this can be desirable for preventing the edge gradients from reaching the peeling-ballooning limit (as discussed in section \ref{subsec:alphat}). At higher values this increased transport can lead to a complete collapse of the plasma, triggering a density limit disruption. In Rogers, Drake \& Zeiler, 1998 \cite{Rogers1998-xp}, it was proposed that the density limit is triggered at low $\beta$, where interchange turbulence is no longer damped due to electromagnetic induction but rather enhanced by electromagnetic transport. The condition for the density limit is given in terms of the ratio of electromagnetic induction to electron inertia (as given in \ref{eq:kEM}), and of the wavenumber of resistive ballooning mode. If the following condition is fulfilled in L-mode (i.e. the condition in \ref{eq:LH} is not met), we expect that an operational point will undergo a density-limit disruption;
\begin{align}
    k_{EM} > \sqrt{\frac{\alpha_c}{\alpha_t} \sqrt{\frac{2\lambda_{p,e,L}}{R}}}\label{eq:LDL}
\end{align}
Here, $\lambda_{p,e,L}$ is the electron pressure gradient near the separatrix for \textit{L-modes} specifically — given in terms of the sound Larmor radius $\rho_{s0}$ and $\alpha_t$ in equation B.1 of ref \cite{Manz2023-vo} — rather than for \textit{H-modes} as in section \ref{subsec:LH}. This is because, so long as the turbulence suppression condition given by equation \ref{eq:LH} is met, catastrophic transport conditions cannot be reached. Instead, within the SepOS framework, H-mode density limits are described as a preceded by a HL back-transition and then an L-mode density limit — as we show in section \ref{sec:dedicated_discharges}.
We can expand and solve equation \ref{eq:LDL} for $n_{e,sep}$, giving
\begin{equation}\label{eq:ldlgreenwald}
    n_{e,sep} > n_{GW} \times 0.11\cdot\frac{\sqrt{\alpha_c}}{\hat{\kappa}^2}\sqrt{\frac{T_{e,sep}}{Z_{eff,sep}}}\lambda_{p,e,L}^{\frac{1}{4}}R^{\frac{1}{4}}
\end{equation}
where $n_{GW}$ is the density limit given in Greenwald et al., 1998 \cite{Greenwald1988-gn} and $\hat{\kappa}=\sqrt{\frac{1}{2}\left(1 + \kappa_{sep}^2 + \left( 1 + 2 \delta_{sep}^2 - 1.2 \delta_{sep}^3 \right)\right)}$. Here, we see that the density limit proposed here is similar to the Greenwald density limit, with a slight additional dependence on the separatrix power (which sets $T_{e,sep}$ via equation \ref{eq:Te_sep}). For a more detailed look at the L-mode density limit, see Manz, Eich \& Grover et al., 2023 \cite{Manz2023-vo}.

\subsection{The ideal MHD limit}\label{subsec:MHD}

At sufficiently high $\beta$, the discharges are limited by the ideal MHD limit. If the following condition is fulfilled, the transport will increase until this condition is recovered;
\begin{align}
    \alpha_{MHD}=\frac{R \, \hat{q}^2_{cyl}}{\lambda_{p,e,H}} \, \beta_{e,sep}<\alpha_c \label{eq:MHD}
\end{align}
for $\alpha_c$ defined in equation \ref{eq:alpha_c}.
 
\section{Demonstrating regime boundaries in dedicated ASDEX Upgrade discharges}\label{sec:dedicated_discharges}

\begin{figure}
    \centering
    \includegraphics[width=1\linewidth]{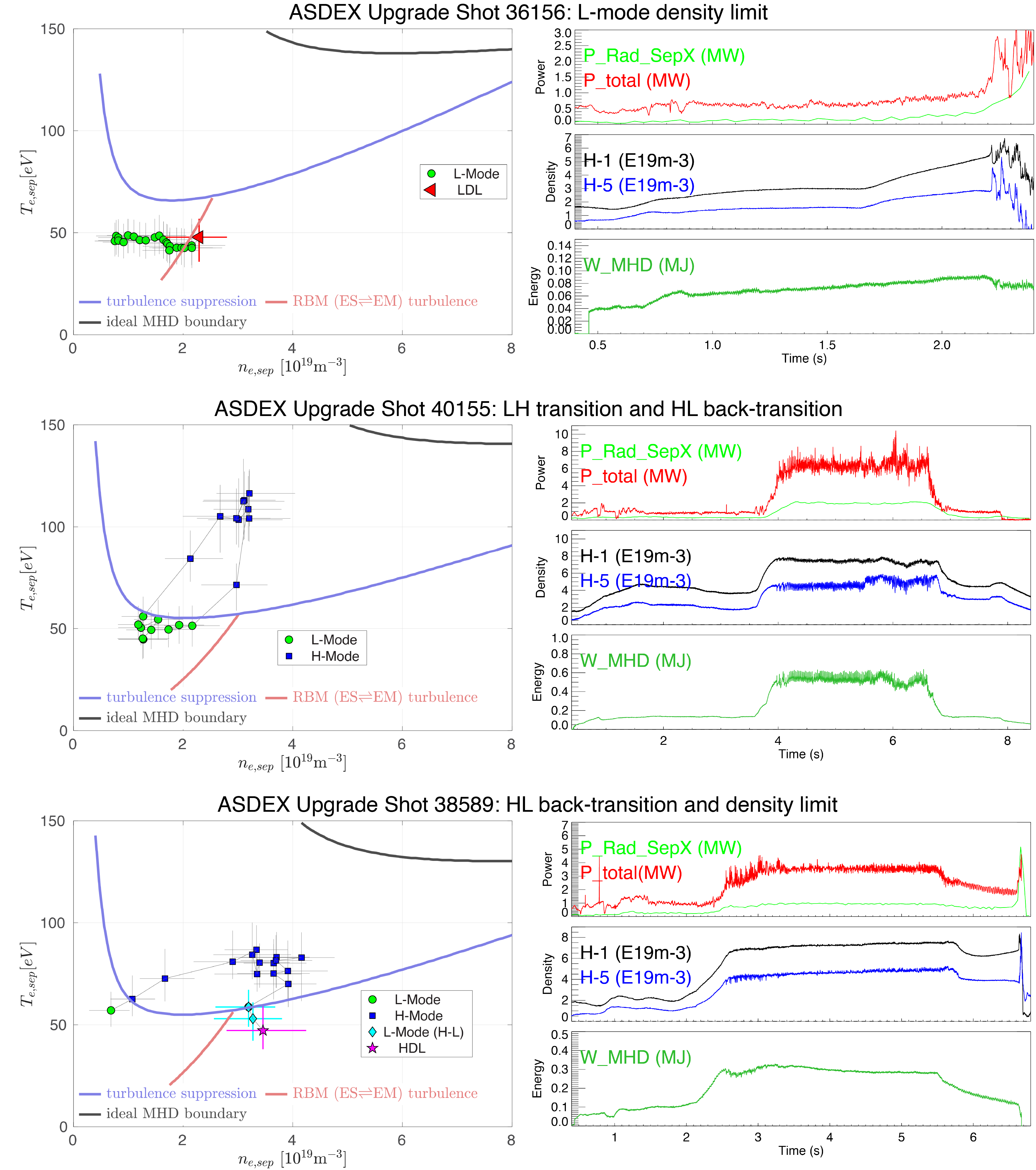}
    \caption{Separatrix measurements (\textit{left column}) and time-traces (\textit{right column}) from three dedicated ASDEX upgrade discharges, showing an L-mode reaching the density limit (\textit{top row}), a LH transition followed by a stable HL back-transition (\textit{center row}) and a LH transition followed by a disruptive HL back-transition (\textit{bottom row}). In the \textit{left column}, the separatrix operational space shows Thomson measurements averaged over $\sim200-400ms$, with consecutive measurements connected via a line to indicate the time-evolution of the discharge. In the \textit{right column}, the time-traces show the total heating power (\textit{red line, top panel}) and the tomographically-constructed confined-region radiation (excluding the X-point region,  \textit{green line, top panel}) \cite{David2021-pa}, the core and edge densities (\textit{black and blue lines respectively, center panel}), and the MHD stored energy (\textit{green line, bottom panel}) for the discharges.}
    \label{fig:dedicated_discharges}
\end{figure}

To demonstrate how to interpret the separatrix operational space and as a step towards its validation, we performed a series of ASDEX Upgrade dedicated discharges. Three of these discharges are shown in Fig. \ref{fig:dedicated_discharges}, showing a discharge designed to probe the L-mode density limit boundary (\textit{top row}), another to demonstrate a LH transition and a stable HL back-transition (\textit{middle row}), and finally a LH transition followed by a disruptive HL back-transition (\textit{bottom row}). 

In the first discharge (\#36156) — shown in the top row of Fig. \ref{fig:dedicated_discharges} — the gas puff is ramped up at constant heating power, steadily increasing the edge and core densities until a density limit disruption is observed. In the SepOS diagram, we see the L-mode points (\textit{green circles}) gradually moving to higher separatrix densities while the separatrix temperature due to the constant heating power remains the same, until the discharge terminates with a density-limit (\textit{red triangle}) at the boundary predicted by condition \ref{eq:LDL}.

In the second discharge (\#40155) — shown in the middle row of Fig. \ref{fig:dedicated_discharges} — the gas puff level is kept constant and the heating power is increased by several MW at 4 s, resulting in a clear transition to H-mode as seen by the sharp increase in the confined-region density and stored energy. At 6.5 s the heating power is reduced to the original level and the plasma transitions back into L-mode. In the SepOS diagram, the increase in heating power is seen as an increase of the separatrix temperature, as we expect from equation \ref{eq:Te_sep}. Despite constant gas-puffing, we see a significant change in $n_{e,sep}$ during the LH transition as the particle confinement improves. Interestingly, after the HL back transition, the L-mode remains at a slightly higher separatrix density than from the point where it transitioned into H-mode.

Finally, in the third discharge (\#38589) — shown in the bottom row of Fig. \ref{fig:dedicated_discharges} — a LH transition is induced by a step in the heating power early in the discharge while the fuelling is continuously increased. Then, at $5.5s$, the heating power is reduced to its original level — similar to in \#40155. However, this time the density of the L-mode after the HL back-transition is above the L-mode density limit, and as a consequence the L-mode disrupts. In the SepOS diagram, we see that this disruption occurs at higher density than the L-mode density limit seen in \#36156. These points are not directly reachable from L-mode because they would disrupt already at lower densities, and as such we label these points as corresponding to a H-mode density limit (HDL). Despite the name, these density-limit disruptions do not occur in H-mode directly, but rather first with a HL back-transition and a subsequent disruption of the over dense L-mode. In the database considered, we have not identified any density-limit disruptions with H-mode separatrix conditions (i.e. fulfilling condition \ref{eq:LH}).

\section[The effect of varying Ip and Bt on the separatrix operational space]{The effect of varying $I_p$ and $B_t$ on the separatrix operational space}\label{sec:varying_parameters}

In the original SepOS paper \cite{Eich2021-df} and in figure \ref{fig:sepos_aug_0.8MA_2.5T}, we showed the separatrix operational space determined for a fixed plasma current of $I_p=0.8\,\mathrm{MA}$ and fixed magnetic field strength of $2.5\,\mathrm{T}$, with the experimental data points selected for a narrow range around these values.

\begin{table}[htb]
\centering
\begin{tabular}{ccccccc}
I$_{p}$& B$_{tor}$&$\hat{q}_{cyl}$&P$_{sep}$&P$_{heat}$&f$_{GW}$& $\alpha_{c}$\\
\hline
[MA]&[T]&-&[MW]&[MW]&-&-\\
\hline
0.4-1.2&1.5-3.0&2.3-8.3&0.3-14&0.4-18&0.2-1.0&2.1-3.2\\
\hline
\end{tabular}
\caption{Discharge parameters of analyzed ASDEX Upgrade L-mode and H-mode plasmas.}
\label{tab:fulldatabase}
\end{table}

To apply the separatrix operational space in a predictive way, we need to show that it is valid as the magnetic field strength and plasma current is varied. We developed an extended database with 6688 individual measurements collected across 524 ASDEX Upgrade discharges over the range of conditions given in table \ref{tab:fulldatabase}. For each measurement point, we used the SepOS framework to calculate whether a point should be in L-mode, H-mode, or density-limit disrupting, and then compared this to the actual state of the plasma during the measurement.

\begin{figure}[tbh]
    \centering
    \includegraphics[width=0.7\textwidth]{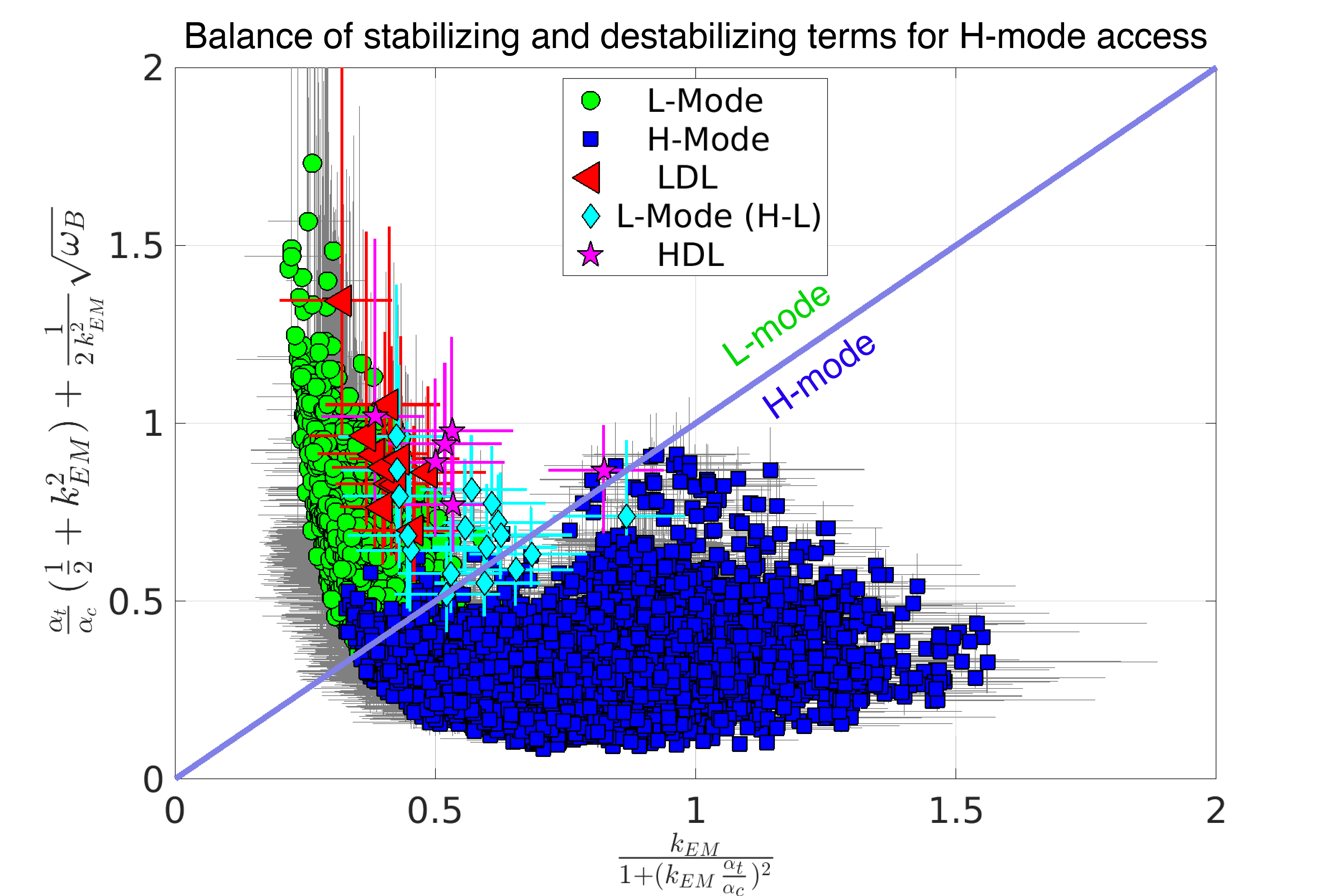}
    \caption{The balance of destabilizing and stabilizing terms (left and right-hand-sides of condition \ref{eq:LH} respectively), for all points in the expanded ASDEX Upgrade database given in table \ref{tab:fulldatabase}. The x-axis gives the rate at which energy flow into the shear flow (stabilizing), and the y-axis gives the rate at which turbulent energy will be generated (destabilizing). Points with stronger stabilization than destabilization are expected to be in H-mode, and vice-versa, with the separation indicated by the $y=x$ line. The actual state of the plasma during the measurement is indicated by the color of the point.}
    \label{fig:LH_full_database}
\end{figure}

In figure \ref{fig:LH_full_database}, we show the balance of the stabilizing (\textit{left-hand-side}) and destabilizing (\textit{right-hand-side}) terms of condition \ref{eq:LH} as the $x$ and $y$ axes respectively — equivalent to figure 3 of ref \cite{Eich2021-df} with the expanded database. For points where the stabilization due to the shear flow exceeds the rate of turbulent energy production we expect the plasma will be in H-mode, and vice-versa. A perfect prediction by SepOS would have all H-mode points (\textit{blue squares}) in the lower-right triangle, and all other points in the upper-right triangle, with the $y=x$ line (\textit{solid blue line}) perfectly separating the two. As we see in figure \ref{fig:LH_full_database}, the agreement is remarkably good — a few H-mode points with weak stabilizing and destabilizing terms (i.e. lower left corner) are not described as accurately, but otherwise the method is able to summarize the entire dataset with an impressive combination of accuracy and simplicity.

Dimensionless comparisons such as figure \ref{fig:LH_full_database} are able to treat databases gathered across different machine conditions ($I_p$, $B_t$, etc) and even different machines, since these dimensionless quantities can be computed for each individual point. However, while this is useful for testing the theory, it is difficult to build an intuitive interpretation of these results.

\begin{figure}
    \centering
    \includegraphics[width=0.69\linewidth]{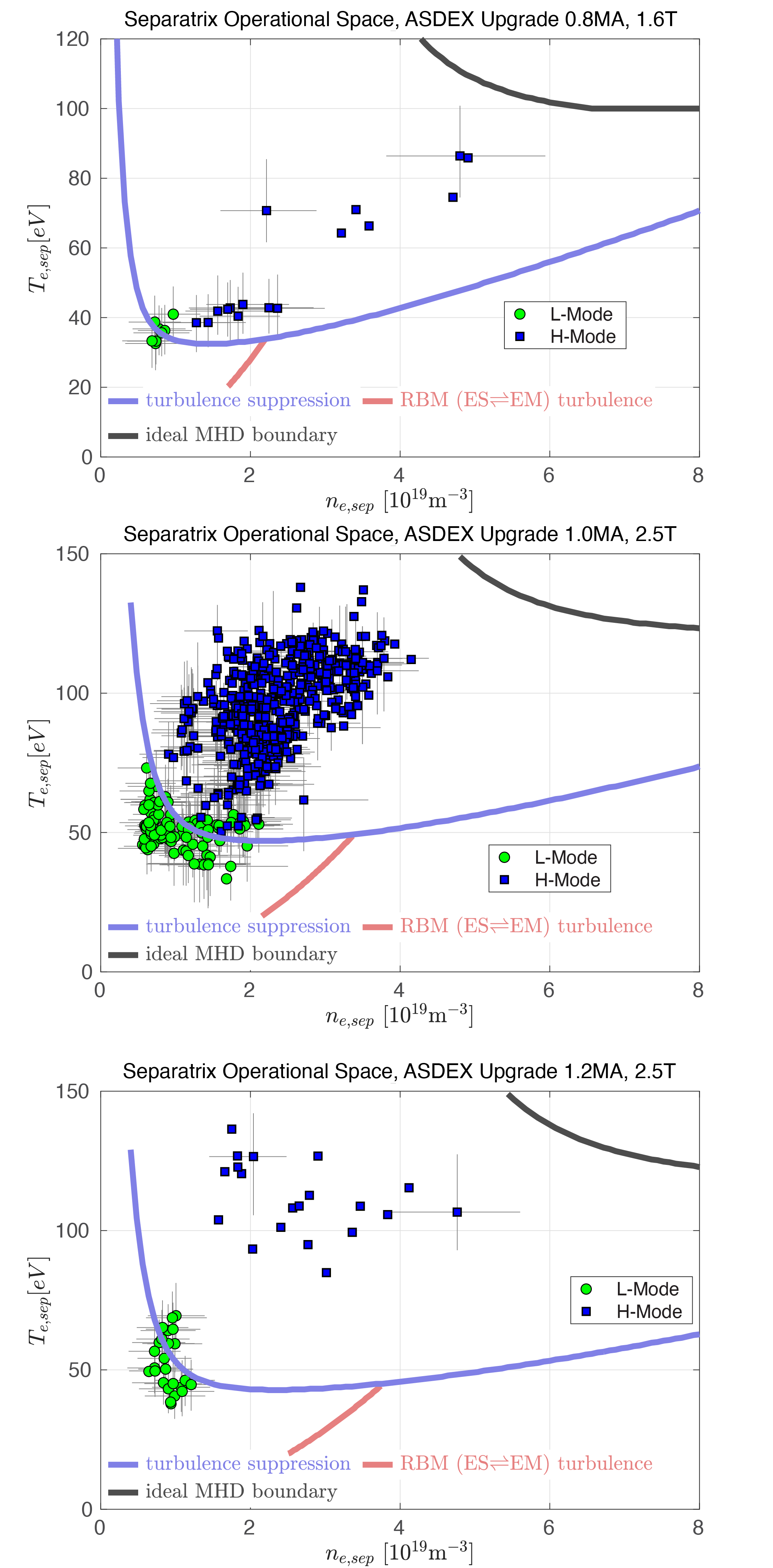}
    \caption{The effect of varying the plasma current and toroidal magnetic field on ASDEX Upgrade, showing the separatrix operational space and Thomson measurements for $I_p=0.8\,\mathrm{MA}, B_t = 1.6\,\mathrm{T}$ (\textit{top row}), $I_p=1.0\,\mathrm{MA}, B_t = 2.5\,\mathrm{T}$ (\textit{middle row}) and $I_p=1.2\,\mathrm{MA}, B_t = 2.5\,\mathrm{T}$ (\textit{bottom row}).}
    \label{fig:sepos_other_conditions}
\end{figure}

To recast the above comparison in terms of separatrix parameters, as in figure \ref{fig:sepos_aug_0.8MA_2.5T}, we need to restrict the experimental database to narrow ranges of machine parameters and determine the dimensional SepOS for those conditions. This is because the SepOS equations depend on quantities such as $\hat{q}_{cyl}$, which depend on machine parameters like the plasma current and magnetic field strength. As seen in figure \ref{fig:sepos_other_conditions}, this means that when we show the results in terms of $n_{e,sep}$ and $T_{e,sep}$, the SepOS transitions change as functions of $I_p$ and $B_t$. 

In figure \ref{fig:sepos_other_conditions} we see that the H-mode access (\textit{blue line}) moves to lower $T_{e,sep}$ as the plasma current is increased and as the magnetic field is decreased. Since $P_{sep,e} \propto T_{e,sep}^{7/2}$ (from equation \ref{eq:Te_sep}), this corresponds to a decrease in the required power crossing the separatrix (in the electron channel) with increasing plasma current or decreasing magnetic field strength. This matches the experimental measurements well, as well as the expected increase in the LH threshold with the magnetic field strength.
However, the result seems to contradict the general observation that the LH power threshold for the high density branch does not have an explicit current dependence \cite{Martin2008-vw}. We propose several resolutions for this apparent contradiction. Firstly, the plasma current affects the heat flux decay length $\lambda_q \propto 1/I_p$ (at fixed machine parameters). Because of this, as $I_p$ is increased less $P_{sep,e}$ is required to maintain the same $T_{e,sep}$. This will at least partly counteract the dependence of $T_{e,sep}$ on $I_p$, resulting in a weaker dependence of $P_{sep}$ on $I_p$ and bring our results closer to ref \cite{Martin2008-vw}. A second point to consider is that we have only discussed the scaling of $P_{sep,e}$, ignoring the contribution of the ion heat flux which is likely dominant \cite{Ryter2014-jb,Schmidtmayr2018-ga}. In section \ref{sec:future_work}, we propose future work which may reduce this uncertainty, but at this point we simply note that the results here should not be over-interpreted. Finally, there are other machine parameters beyond $I_p$ and $B_t$ which can affect the SepOS results. These include the shaping, which strongly affects $\alpha_c$, and the machine dimensions, which requires validation on other machines as discussed in section \ref{sec:future_work}.

\section{The normalized separatrix operational space}\label{sec:normalized_sepos}

In the previous section, we showed two approaches to generalize the SepOS framework for changing machine parameters. The first — in terms of normalized variables — worked for arbitrary machine parameters but was non-intuitive, while the second — in terms of separatrix variables for slices of the database — had the opposite problem. Is it possible to somehow combine the two approaches, to get the best of both? This led us to normalize the separatrix variables to the LH-minimum density and temperature — which gave some surprising results.

\begin{figure}
    \centering
    \includegraphics[width=0.7\linewidth]{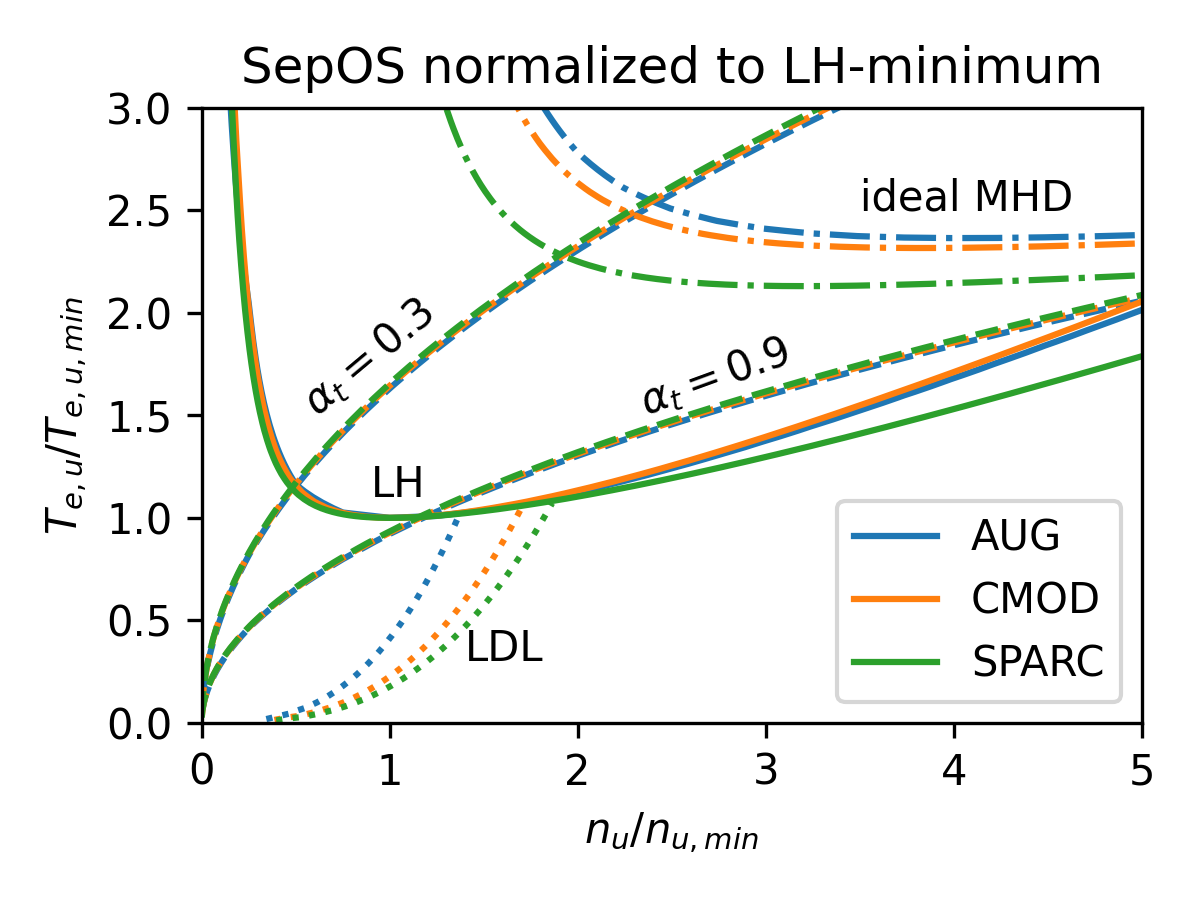}
    \caption{SepOS for ASDEX Upgrade, Alcator C-Mod and SPARC, normalized to their respective LH-minima given in table \ref{tab:normalized_sepos_parameters}. The normalized LH transition, $\alpha_t$ contours, L-mode density limit and ideal MHD limit are shown for each machine, where the line color indicates the machine shown and the line style indicates which limit is being shown (\textit{solid line} for LH transition, {dashed line} for $\alpha_t$ contours, \textit{dotted line} for L-mode density limit and \textit{dot-dashed line} for ideal MHD).}
    \label{fig:normalized_sepos}
\end{figure}
\begin{table}[h]
    \centering
\begin{tabular}{|c|c|c|c|}
\hline
&ASDEX Upgrade&Alcator C-Mod&SPARC\\
\hline
$B_0$ [T]&2.5&5.4&12.2\\
\hline
$R_0$ [m]&1.65&0.7&1.85\\
\hline
$a$ [m]&0.49&0.25&0.57\\
\hline
$I_p$ [MA]&0.8&1.0&8.7\\\hline
 min $n_{e,sep}$ [$n_{19}$]& 2.0& 7.6&16.6\\\hline
 min $T_{e,sep}$ [$eV$]& 55& 72&88\\\hline
    \end{tabular}
    \caption{Machine parameters used for figure \ref{fig:normalized_sepos}. The shaping is held fixed at $\kappa_{95}=1.65, \delta_{95}=0.3$.}
    \label{tab:normalized_sepos_parameters}
\end{table}

In figure \ref{fig:normalized_sepos} we see that the LH transition curves for ASDEX Upgrade, Alcator C-Mod and SPARC more-or-less overlap when normalizing to their LH minima (given in table \ref{tab:normalized_sepos_parameters}). The L-mode density limit and ideal MHD limits are closer than without the normalisation but do not exactly overlap. The use of the ASDEX Upgrade L-mode $\lambda_{p,e,L}$ (equation B1 from ref \cite{Manz2023-vo}) for C-Mod and SPARC is not appropriate, since this scaling was not elaborated to have any explicit magnetic field, plasma current or machine size dependence. Nevertheless, it is interesting to consider why the L-mode density limit moves to higher normalized densities on C-Mod and SPARC. As discussed in section \ref{subsec:LDL}, the L-mode density limit scales approximately linearly with the plasma current\cite{Greenwald1988-gn}, while the LH transition does not\cite{Martin2008-vw}, as such operating with a higher plasma current reduces the LH density minimum relative to the density limit in the model used. We remind the reader that the actual decay length only enters with the fourth root in equation \eqref{eq:ldlgreenwald}.

We also see that the contours of $\alpha_t$ overlay exactly once normalized to the LH minima. This result is noteworthy as it has an important implication. As we discussed in section \ref{subsec:alphat}, the ELM-free QCE regime is found for $\alpha_t > 0.55$ on ASDEX Upgrade \cite{Faitsch2023-ws}. With this result, that suggests that machines operating with densities above the LH minimum density and sufficient shaping should generally be able to access this regime.

In figure \ref{fig:normalized_sepos}, we normalized the SepOS results from each machine according to the computed LH contour. How does this point change when considering different machines? We should, in principle, be able to find the point where the derivative of equation \ref{eq:LH} with respect to density equals zero, indicating the minimum density. However, this is not straightforward because $\lambda_{p,e}$ depends on $\alpha_t$, and so we instead perform a regression on the model results. We evaluated the SepOS for 19 sets of parameters representing a range of C-Mod, ASDEX Upgrade, JET, ITER and SPARC conditions, for $B_t$ ranging from $1.47 - 12.2\,\mathrm{T}$, $I_p$ ranging from $0.62 - 15\,\mathrm{MA}$, minor radius $a$ from $0.25 - 2.0\,\mathrm{m}$ and major radius $R$ from $0.7 - 6.2\,\mathrm{m}$. The best fit for the minimum density across these cases is
\begin{align}
n_{e,min}[10^{19}\,\mathrm{m}^{-3}]=(0.55\pm 0.07) B_T^{0.81\pm 0.02} I_p^{0.36\pm 0.02} a^{-0.89\pm 0.03}\label{eq:mindensshift}
\end{align}
for $B_T$ in T, $I_p$ in MA and $a$ in m. This can be compared to equation 3 from Ryter et al., 2014 \cite{Ryter2014-jb}, setting the aspect ratio for ASDEX Upgrade to $R/a=1.65\,\mathrm{m}/0.5\,\mathrm{m}=3.3$ to find
\begin{align}
\bar n_{e,min} [10^{19}\,\mathrm{m}^{-3}] &= 0.7 B_T^{0.62} I_p^{0.34} a^{-0.95} \left(3.3\right)^{0.4}\\
&=1.1 B_T^{0.62} I_p^{0.34} a^{-0.95} \label{eq:Ryter_scaling_simplified}
\end{align}

These scalings agree remarkably well, considering that they were determined via entirely different methods. The most significant difference is that of the prefactor, which is a factor of $2$ higher in equation \ref{eq:Ryter_scaling_simplified}. However, our scaling is in terms of the separatrix density $n_{e,sep}$ while equation \ref{eq:Ryter_scaling_simplified} is in terms of the line-average density $\bar n_e$, which is typically $2-3\times$ higher than $n_{e,sep}$ in H-modes. This difference is enough to explain the difference in prefactor, although with a more typical $\bar n_e / n_{e,sep} = 3$ we are actually predicting a $50\%$ higher minimum density than Ryter et al., 2014 \cite{Ryter2014-jb}. Similar to our discussion in section \ref{sec:varying_parameters}, there are multiple suspects for this remaining difference, chiefly the ion heat flux, which we will discuss in section \ref{sec:future_work}. Nevertheless, even before this remaining factor is identified, the agreement between equations \ref{eq:mindensshift} and \ref{eq:Ryter_scaling_simplified} suggests that the SepOS can be used to inform the operations of other machines, including next-step devices. 

\section{Predicting QCE access on SPARC using the separatrix operational space}\label{sec:qce_access_on_sparc}

\begin{figure}
    \centering
    \includegraphics[width=0.7\linewidth]{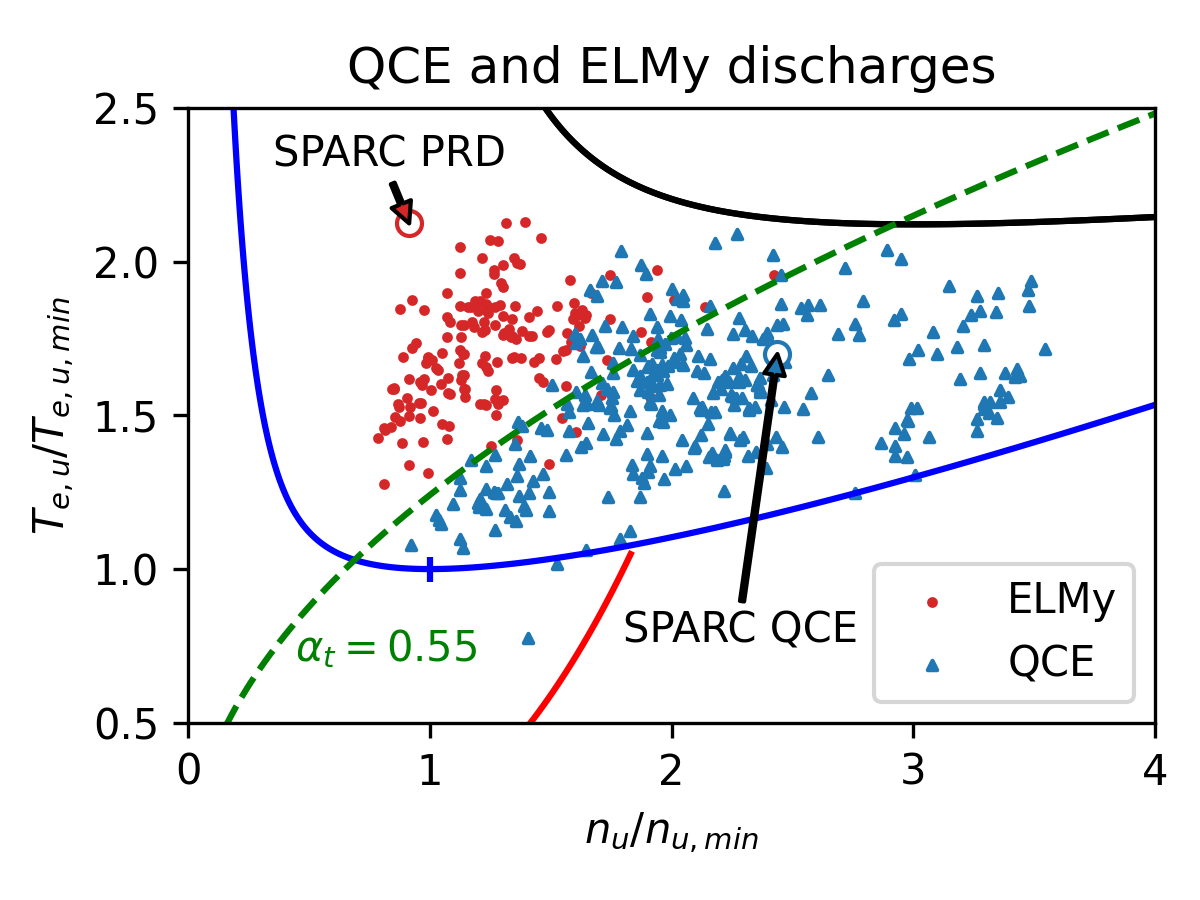}
    \caption{QCE access in the normalized separatrix operational space, using the ASDEX Upgrade database from Faitsch et al., 2023 \cite{Faitsch2023-ws}. Type-I ELMy points are indicated with \textit{red circles}, and QCE points are indicated with \textit{blue triangles}. The normalized LH transition is shown as a \textit{solid blue line}, and the $\alpha_t = 0.55$ (QCE/ELMy transition) is shown as a \textit{green dashed line}. The SPARC PRD (primary reference discharge) and a proposed SPARC QCE operational point are indicated at the annotated points.}
    \label{fig:qce_on_norm_sepos}
\end{figure}

\begin{table}[htb]
    \centering
    \begin{tabular}{|c|c|c|}
\hline
&PRD&QCE\\
\hline
$\alpha_t$&0.2&0.7\\
\hline
n$_{e,sep}$&1.5E20\,m$^{-3}$&4.0E20\,m$^{-3}$\\
\hline
T$_{e,sep}$&195\,eV&156\,eV\\
\hline
$\frac{n_{e,sep}}{n_{GW}}$&0.17&0.45\\
    \hline
    \end{tabular}
    \caption{Separatrix values for the PRD (type-I ELMy H-Mode) and QCE discharge in SPARC for I$_p$\,=\,8.7\,MA, B$_t$\,=\,12.2\,T, $P_{sep}$\,=\,20\,MW, n$_{GW}$\,=\,8.7E20\,$\mathrm{m}^{-3}$.}
    \label{tab:prdqce}
\end{table}

What does the separatrix operational space look like for a next-step device like SPARC? 
SPARC's primary reference discharge (PRD) is a high-performance $Q\sim 11$, full-field $B_t \sim 12.2\,\mathrm{T}$, full-current $I_p \sim 8.7\,\mathrm{MA}$ H-mode \cite{Creely2020-yk}. This operating point is predicted to have a separatrix density of $n_{e,sep}\sim 1.5\times10^{20}\,\mathrm{m}^{-3}$ and a separatrix power of $P_{sep}\sim20\,\mathrm{MW}$, corresponding to a separatrix temperature of $T_{e,sep}\sim 195\,\mathrm{eV}$. Normalizing these values to the LH minimum of $1.64\times10^{20}\,\mathrm{m}^{-3}, 91.8\,\mathrm{eV}$ and comparing to figure \ref{fig:qce_on_norm_sepos}, we see that this puts the PRD operational point squarely in the type-I ELMy regime. To deal with this, SPARC has the ability to use its error field correction coils as a resonant magnetic perturbation (RMP) ELM-suppression system. Nevertheless, intrinsically ELM-free scenarios are attractive, especially for scaling to a power plant. 

With the SepOS we identified a broad operational space on SPARC that we expect will be in the QCE regime. Within this space we propose a point maintaining the PRD's separatrix power of $P_{sep}\sim20\,\mathrm{MW}$, but with a significantly higher separatrix density of $n_{e,sep}\sim4\times10^{20}\,\mathrm{m}^{-3}$. The separatrix temperature is also slightly lower, at $T_{e,sep}\sim156\,\mathrm{eV}$, due to $\lambda_q$ broadening at increased $\alpha_t$ \cite{Eich2020-jx}. This point is also more attractive for steady-state heat-exhaust, since the increased separatrix density significantly reduces the edge impurity concentration required for detachment. As part of ongoing work, this QCE operational space will be investigated in \href{https://github.com/cfs-energy/cfspopcon}{cfspopcon} to determine the highest-gain point compatible with detached, ELM-free conditions in the edge.

\section{Conclusions and future work}\label{sec:future_work}

The separatrix operational space (SepOS) provides a simple, fast and intuitive framework for ensuring benign heat exhaust conditions when designing future devices and planning their operation. The SepOS framework is relatively simple, expressed concisely in terms of separatrix parameters, separatrix gradients and machine parameters. Rather than a formal derivation, the framework selects terms from linear turbulence models to develop an accurate description of an extensive database of ASDEX Upgrade experimental results. Perhaps unsurprisingly, this model provides an excellent description of the experimental database with which it was derived. However, due to the data-driven approach, before we can use the framework predictively, we needed to rigorously validate it under other conditions — which is the focus of this work.

In this paper, we showed that the SepOS framework remains accurate as the plasma current and magnetic field strength is varied on ASDEX Upgrade. For this, we presented results from an extended database of 6688 individual measurements collected across 524 ASDEX Upgrade discharges, including several discharges which were specifically designed to probe the transitions predicted by SepOS. Across this extended database, the SepOS framework was able to accurately predict whether a given operational point would be in L-mode, H-mode or undergoing a density-limit disruption.
A second, indirect validation was given when realizing that the density of the LH minimum point scaled the same way as in the Ryter scaling \cite{Ryter2014-jb}. Since that scaling broadly agrees with published minima data from devices as diverse as ASDEX Upgrade, C-Mod, DIII-D and JET, finding agreement with that scaling suggests that the SepOS framework (or at least the predicted LH transition) is applicable across multiple devices.

Building on these results, we then demonstrated how the SepOS framework can be applied predictively, to identify an operational point expected to be in the QCE regime for SPARC. This proposed operational point will be investigated in more detail as an attractive operational point for combining high fusion gain with relatively benign heat-exhaust conditions. Once SPARC starts H-mode operations in its second campaign, testing whether the SepOS framework predicts the access to the QCE regime will be a strong validation (or falsification) of the model and its use for fusion power plants.\\

Rather than a conclusive validation of the SepOS framework, this work should be considered a starting point and a call to action. Future work is needed to investigate strongly-seeded discharges, to determine how the SepOS predictions are modified as plasmas are pushed towards the detached conditions needed for tolerable steady-state heat exhaust. A further step would be a detailed investigation into the ion temperature dynamics, which would be particularly important for translating the results from a separatrix temperature to power fluxes across the separatrix. Similar to their use in Grover et al., 2024 \cite{Grover2024-hd}, this could be a useful application of high-fidelity turbulence models such as GRILLIX \cite{Zholobenko2024-df}. Further investigations should probe the underlying physics in different regions of the separatrix operational space, as a rigorous check against fortuitous agreement. Finally, the model must be applied to other, existing devices. A validation of the model against C-Mod data is in progress and a publication is in preparation.  By performing a multi-machine validation of SepOS across various devices such as JET, TCV, MAST-U, and DIII-D, we can test and improve the predictive capabilities of the framework, letting us confidently apply it for the design of future fusion power plants with optimized heat exhaust solutions.

\section{Acknowledgements}
This work was supported by Commonwealth Fusion Systems.
This work has been carried out within the framework of the EUROfusion Consortium, funded by the European Union via the Euratom Research and Training Programme (Grant Agreement No 101052200 — EUROfusion). Views and opinions expressed are however those of the author(s) only and do not necessarily reflect those of the European Union or the European Commission. Neither the European Union nor the European Commission can be held responsible for them.

\printbibliography

\end{document}